# High quality waveguides for the mid-infrared wavelength range in a silicon-on-sapphire platform

Fangxin Li,[1] Stuart D. Jackson,[1] Christian Grillet,[1] Eric Magi,[1] Darren Hudson,[1]
Steven J. Madden,[2] Yashodhan Moghe,[3] Christopher O'Brien,[3*] Andrew Read,[3]
Steven G. Duvall,[3] Peter Atanackovic,[3] Benjamin J. Eggleton,[1] and David J. Moss[1**]

[1] Centre for Ultrahigh-bandwidth Devices for Optical Systems (CUDOS), Institute of Photonics and Optical Science (IPOS), School of Physics, NSW 2006, Australia
[2] CUDOS, Australian National University, Canberra, ACT 2006, Australia
[3] Silanna Semiconductor, 8 Herb Elliot St., Sydney Olympic Park, New South Wales 2137, Australia
[*] Present Address: Peregrin Semiconductor, California USA
[**] Corresponding author: dmoss@physics.usyd.edu.au

**Abstract:** We report record low loss silicon-on-sapphire nanowires for applications to mid infrared optics. We achieve propagation losses as low as 0.8dB/cm at λ=1550nm, ~ 1.1 to 1.4dB/cm at λ=2080nm and < 2dB/cm at λ = 5.18 μm.

1. Introduction

Photonic integrated circuits for the mid-infrared (2-20μm) wavelength range is a new and burgeoning area of interest [1-8] for a variety of fundamental applications, including thermal imaging (2.5 - 15μm) [9], chemical bond spectroscopy (from the visible to 20 μm and beyond), astronomy [10], gas sensing, and military applications such as missile countermeasures. Historically, however, the mid-infrared has posed challenges for photonics. Coherent mid-infrared laser sources have been bulky and expensive, or have required cryogenic cooling, as did common mid-infrared detectors. Recently, inexpensive and reliable single-mode quantum cascade lasers have become commercially available for wavelengths as long as 9 μm, and with powers from 10 - 100 mW, and this has begun to change the landscape [11]. In addition, single-mode fibers are now available at wavelengths out to 6 μm [12,13], as are mid-infrared photodetectors with bandwidths over 1 GHz [6]. As a result, building a single-mode optical system in the mid-infrared is now within both financial and technical reach. However, the lack of a suitable platform for integrated optical waveguides at these long wavelengths has meant that to date, mid-infrared systems have been largely implemented with free-space optics.

Silicon, in the form of silicon-on-insulator (SOI) waveguides, has attracted significant interest recently as a potential platform for integrated optical devices for the mid-IR [1-6,]. Silicon itself is transparent out to beyond 6μm, and has the important benefit that its nonlinear properties are actually much better in the mid-IR than they are in the near IR [3,4]. Beyond ~ 2μm two-photon absorption drops to zero, and so the well known problem of TPA and the associated free carriers becomes negligible. This has motivated some recent demonstrations of nonlinear optics in the mid IR in SOI waveguides, near 2μm [3,4]. However, for applications at longer wavelengths, the silica cladding layer that is used in the SOI platform becomes more strongly absorbing, and substrate leakage can become an issue [5] and so it is clearly of interest to explore other platforms. While exotic structures such as hollow-core and freestanding waveguides have been proposed in order to avoid these problems [13-15], probably the greatest potential lies in finding a new substrate platform that is intrinsically more suited to the mid IR than SOI. Silicon-on-sapphire (SOS) has recently [7,8] attracted interest for mid IR applications for a number of reasons. Not only is sapphire transparent beyond 6μm, but the ability to realise silicon waveguides and nanowires on a low index substrate eliminates any potential substrate leakage, in contrast with SOI where the thin (on the order of μm) silica cladding layer is prone to leakage to the silicon substrate[5]. SOS has the ability to realise high confinement, fully etched waveguides for use from λ=1.2μm to almost 7μm – encompassing both telecommunications and mid-IR wavelengths, while maintaining electronic compatibility.

Here, we demonstrate record low loss waveguides in SOS nanowires. We achieve propagation losses below 1dB/cm in the telecom band near 1550nm, below 1.4dB/cm near λ=2.08μm and less than 2dB/cm at λ=5.18 μm. In particular, our results at λ=5.18 μm represent a significant reduction over the best previously published results in this wavelength range [7,8], and are low enough to offer significant promise for many devices including integrated ring resonators. Our results further support the feasibility of using this platform for the mid-IR, enabling a range of new applications in this wavelength range.

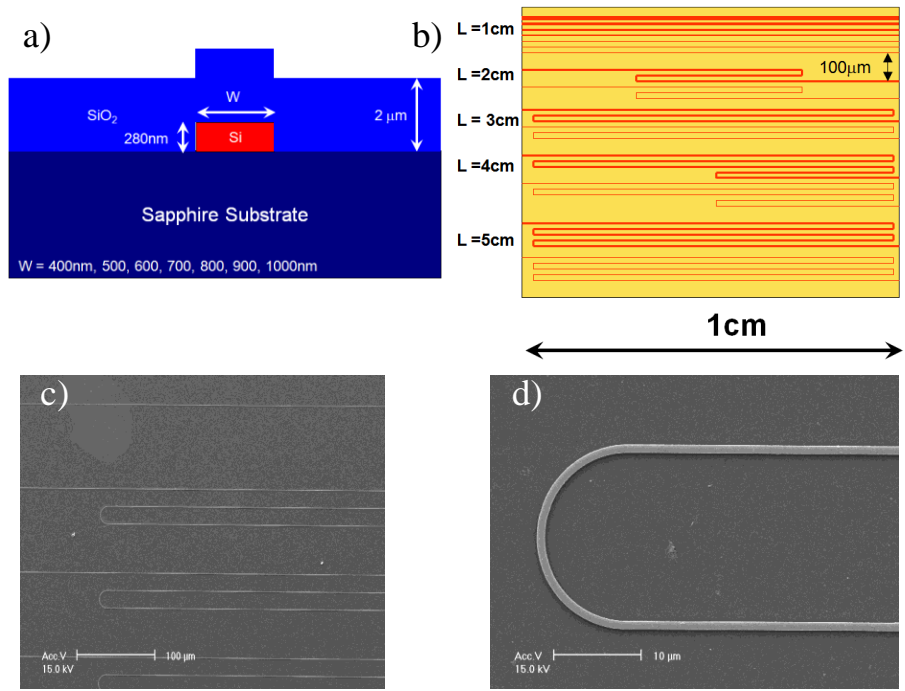

Figure 1. a) Top left: device cross section. Silicon nanowire widths were 400nm, 600nm, 800nm and 1000nm. B)Top right: schematic of 1cm x 1cm chip layout. Note not all devices are shown. c) Bottom left: SEM picture of fabricated device and d) Bottom right: higher magnification image of a bend with a radius of 10μm.

## 2. Experiment

Figure 1 shows a cross section of the device along with a schematic of the chip layout. Epitaxially grown SOS wafers (100 mm diameter, 1mm thick, with 280nm of silicon on top) were patterned with silicon nanowires having lengths of 10, 20, 30, 40 and 48mm and a range of widths of 400, 600, 800 and 1,000 nm 133using an I-line stepper mask aligner, followed by reactive ion dry etching. The wafer was then over-coated with 2 μm of $SiO_2$ using PECVD, and then the final wafer was laser-scribed on the underside and cleaved to 1cm x 1cm chips. Figure 1 summarizes the chip layout (note that all devices are not shown), and also shows SEM images of the final fabricated nanowires. For each length we included the full range of nanowire widths. Waveguides longer than 1cm were accommodated by adding bends (10μm bend radius) with end structures being simply straight nanowires – ie., no attempt was made to optimize coupling. This layout facilitated performing "cut-back" measurements without having to cleave different length samples.

We performed loss measurements at 3 different wavelengths – λ=1550nm, λ=2080nm, and λ=5.18μm, using 3 different experimental setups. At λ=1550nm the measurements were performed using a standard tunable diode laser and coupling was achieved via lensed fiber-tapers controlled with nanopositioning stages, with the ability to image the mode profile on a digital camera.

The measurements at λ=2080nm, were performed using a CW Thulmium doped fiber laser [16] with the same lensed fiber taper coupling setup as the λ=1550nm measurements. The fiber laser output was collimated and coupled into SM2000 fiber that can operate up to 2.1 μm (Thorlabs) containing the lensed taper. The output fiber was directly input to a temperature controlled InGaAs detector (sensitive to 2.4μm). We used a Wollaston prism to polarize the laser output, followed by a polarization paddle controller to control the polarization state.

Figure 2 shows the experimental setup for the measurements at λ=5.18μm. We used a (Daylight Solutions) tunable CW quantum cascade laser (50mW output) capable of producing linearly polarized light from 5.07 μm to 5.37 μm, coupled this into a very short length (50cm) of single mode AsSe chalcogenide fiber [17] having a mode diameter of ~8μm at this wavelength. The laser emitted TE polarized light (horizontal) which was not altered substantially by the short fiber length. We then butt coupled the chalcogenide fiber to the nanowire ensuring the light was TE polarized, and the output was then imaged on a mid IR camera using a ZnSe mid IR objective lens (AR coated) with a focal length of 6mm. The mode profiles were imaged using a Spiricon Beam Profiling Cameras (OPHIR) and averaged 250 times. Relative loss measurements at λ=5.18μm were obtained by measuring the intensity of the central guided mode peak on the imaging camera as a function of waveguide length. This method was highly effective at discriminating against scattered light or substrate guided light. The drawback is that we were only able to obtain relative loss measurements which yielded only propagation loss, not coupling loss.

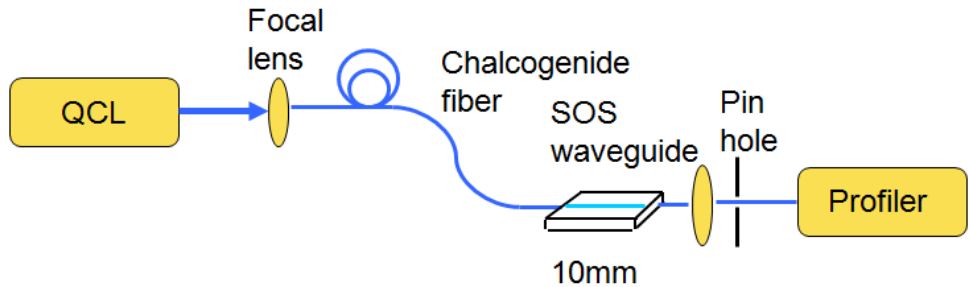

Figure 2. Experimental setup for measurements at λ=5.18 μm. QCL= quantum cascade laser. The chalcogenide fiber was single mode $As_2Se_3$ fiber.

## 3. Results

Figure 3 shows the results of loss measurements at both λ=1550nm and λ=2080nm for both TE and TM polarizations for a range of widths. Note that the absolute scale of the measurements is meaningful here and reflects the taper-taper coupling loss that we were able to achieve with the bare (un-tapered) nanowires. For TE polarized light the losses at λ=1550nm varied from 0.8dB/cm up to 1.2dB/cm. The values below 1dB/cm in particular are remarkably low loss results, considering that the state of the art loss for SOI nanowires was around 3dB/cm for quite some time. Given the low scatter and good linearity of the loss data at λ=1550nm we estimate the experimental uncertainty in our loss measurements is ~

±0.1dB/cm. The variation in loss that we observed from 0.8dB/cm up to 1.2dB/cm did not seem to depend systematically on nanowire width, and instead we ascribe this to random variation in loss across the wafer. The low scatter in the data (relative to its linear variation with length) reflects the high degree of consistency that we were able to achieve in facet quality (and hence facet coupling loss) across the different nanowires for a given cleave. The absolute taper-taper coupling loss was high at ~ 14dB to 16dB / facet but this was expected given the low mode overlap between the tapers (roughly 2 µm beam waist diameter) and the nanowire modes. While the propagation loss did not show a clear cut systematic dependence on nanowire width, the coupling loss did decrease consistently with width, as expected. Figure 3 also shows the cutback measurements at λ=2.08µm using the thulmium doped fiber laser where we achieve losses of 1.1 dB/cm to 1.4dB/cm for TE polarized light and 1.3 dB/cm to 1.7dB/cm for TM.

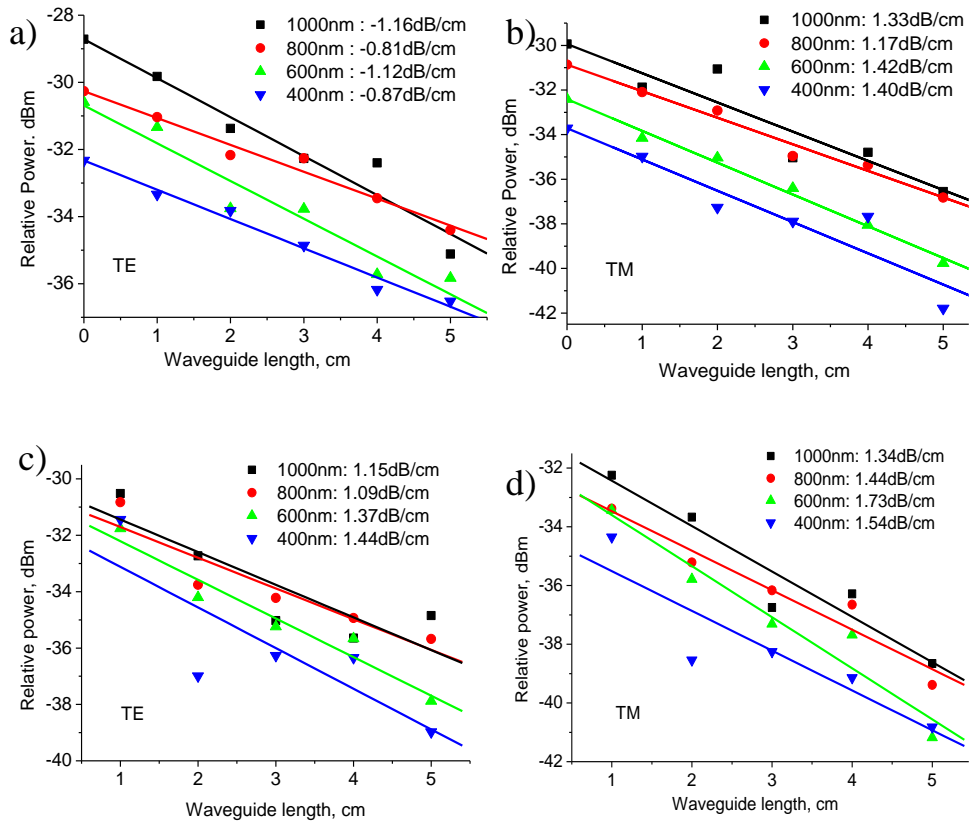

Figure 3. Cutback loss measurements for different nanowire widths (400nm to 1000nm).at λ=1.55 µm for TE (a) Top left) and TM (b) Top right polarizations, and at λ=2.08 µm (c) (Bottom left TE) and d) Bottom right (TM).

Figure 4 shows the results of loss measurements at λ=5.18µm, along with associated mode profiles (TE). We observed very well defined guided modes at λ=5.18µm, and our approach of using the mode field intensity maximum on the imaging camera to measure the relative guided mode intensity versus length, resulted in a very high degree of discrimination from scattered light and yielded very low scatter (deviation from linearity) in the cutback measurements. From Figure 4 we see that the propagation losses were ~1.92±0.05dB/cm for the 1000nm wide nanowires. The error bars in the loss arise from the standard deviation from the linear fit, and are as low as they are because of the low scatter and high linearity of our cutback measurements in Figure 4. The corresponding results for the other widths were ~1.69dB/cm for 800nm wide, 1.85dB/cm for 600nm wide, and 1.90dB/cm for 400nm wide nanowires. We attribute the small variation in propagation loss for different widths to a random variation across the wafer rather than any systematic dependence. The very high quality of the results, including the low scatter and high degree of linearity in the cutback measurements, indicates that the facet loss was extremely consistent from across the wafer. These results represent a reduction by more than a factor of 2 (in dB/cm) over the best results reported in SOS waveguides [7,8], and are particularly remarkable given our relatively thin (280nm) silicon layer.

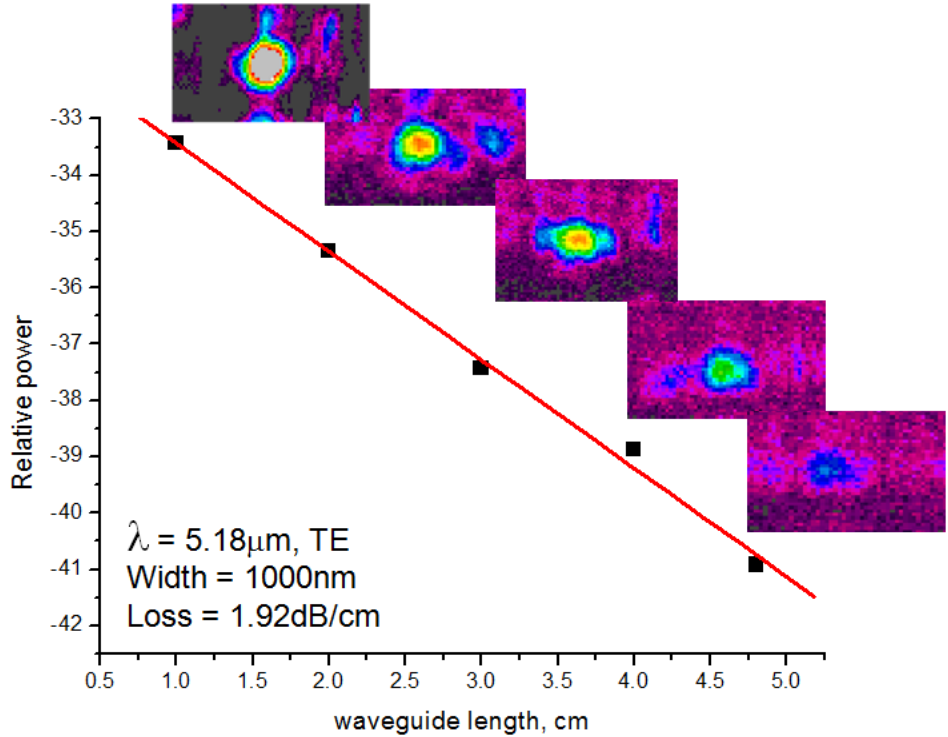

Figure 4. Cutback loss measurements at λ=5.18 μm for W=1000nm (Top) for TE polarization along with imaged mode profiles for each length.

## 4. Discussion

Overall, the propagation losses in our nanowires showed a progressive increase with wavelength in going from λ=1550nm to λ=2080nm, and finally λ=5.18μm, and were generally higher for TM polarization than TE, as expected. The fact that we were able to achieve less than 1dB/cm at λ=1550nm (for TE polarization) is surprising given that typical losses for nanowires in SOI fabricated with electron beam lithography are ~ 3 to 4dB/cm [3,4]. We attribute our low loss to the use of an I-line stepper mask aligner, along with the epitaxial growth process used to produce the SOS wafers, which results in extremely low defect density. The resolution of our stepper mask aligner was ~ 400nm which is much lower than that for typical electron beam lithography machines. We believe this contributed to substantially lower sidewall roughness.

The low propagation loss at λ=5.18μm was unexpected given the thin (280nm) silicon thickness. Figure 5 shows the results of theoretical calculations using a mode solver based on finite element methods (FemSim, RSOFT) for the mode effective index as a function of wavelength, along with mode profiles at two of the wavelengths studied here, for a 1000nm wide nanowire. The mode at λ=5.18μm was actually slightly below cutoff and hence quite delocalized, and so the plot in Figure 5 was obtained by slightly adjusting the experimental parameters (silicon dimensions, PECVD silica cladding refractive index) to the upper end of their experimental range, where the mode was slightly above cutoff. Figure 5 also shows the dispersion curve for this case (green dashed line). This illustrates the extreme sensitivity of the mode near cutoff to experimental error in the device parameters.

Whether or not the modes are near cutoff or slightly below is in fact extremely difficult to determine theoretically for a variety of reasons. Apart from the sensitivity to the experimental parameters discussed above, the accuracy of the calculations was limited by a number of factors, including the extremely small nanowire dimensions (~λ/10), very high index contrast, and the fact that we are interested in the wavelength region near cutoff. We note that even if the mode were slightly below cutoff, however, lossy modes below cutoff have been observed in silicon before [18] with relatively low propagation losses on the order of 1 dB/cm. In this work, it is evident that the contribution to the loss due to the leaky nature of the mode is substantially less than 1dB/cm.

We also expected to observe high loss at λ=5.18μm due to the silica overcladding layer. Figure 6 shows the relative fraction of the mode field energy that resides in the 3 regions (silica cladding, silicon core, sapphire substrate) as a function of wavelength, normalized to the cutoff wavelength. At λ=5.18μm, this corresponds to $\lambda/\lambda_c$ of 1.05 to 1.10, where we see that the fraction of the mode in the upper silica cladding is extremely low, at ~ - 30dB. Based on the absorption of bulk silica at this wavelength, we estimate that the contribution to the overall propagation loss from absorption in the silica cladding is < 1dB/cm – ie., less than the total loss we observed. Figure 6 also shows that at λ=5.18μm the majority of the mode resides in the sapphire substrate. We note that in our structure a cutoff wavelength only exists in the first place because of the asymmetric nature of the waveguides – ie., that the substrate (sapphire) has a slightly higher refractive index than the upper cladding (silica). If not for this, the waveguides would theoretically always have a guided mode. We believe that this is the physical reason why the mode is significantly "pulled" into the substrate near cutoff in our waveguides.

Perhaps the most surprising result is our observation of negligible bendloss at λ=5.18μm. The mask layout was designed such that the number of bends for the different length segments was zero (L=1cm), 2 (L=2cm and 3cm) and 4 (for L=4cm and 4.8cm). Hence, any contribution to the overall loss would produce a deviation from the linear fit of loss versus length. Because of the extremely high quality of our cutback results - very low scatter in the data with high linearity, leading to a very low standard deviation - we are able to attribute an upper limit to the bend loss of 0.1dB / bend in these waveguides. We note that in general bend loss is much more sensitive than the mode index to all of the factors discussed above, particularly near cutoff, and so it is perhaps not surprising that our finite element calculations were not able to predict this. The combination of very high resolution demanded by the small waveguide dimensions with the very extended nature of the mode near cutoff places extreme requirements on the numerical processing. Investigation of these issues using full scale very high resolution 3D FDTD methods on cluster computers will be reported in future work.

Finally, we note that it is not necessarily the intention of this work to advocate these specific device dimensions as being optimal for use near λ= 5 μm. Waveguides designed for use in that wavelength range would ideally have a thicker silicon core, as well as an upper cladding that is intrinsically transparent (such as PECVD deposited alumina for example).

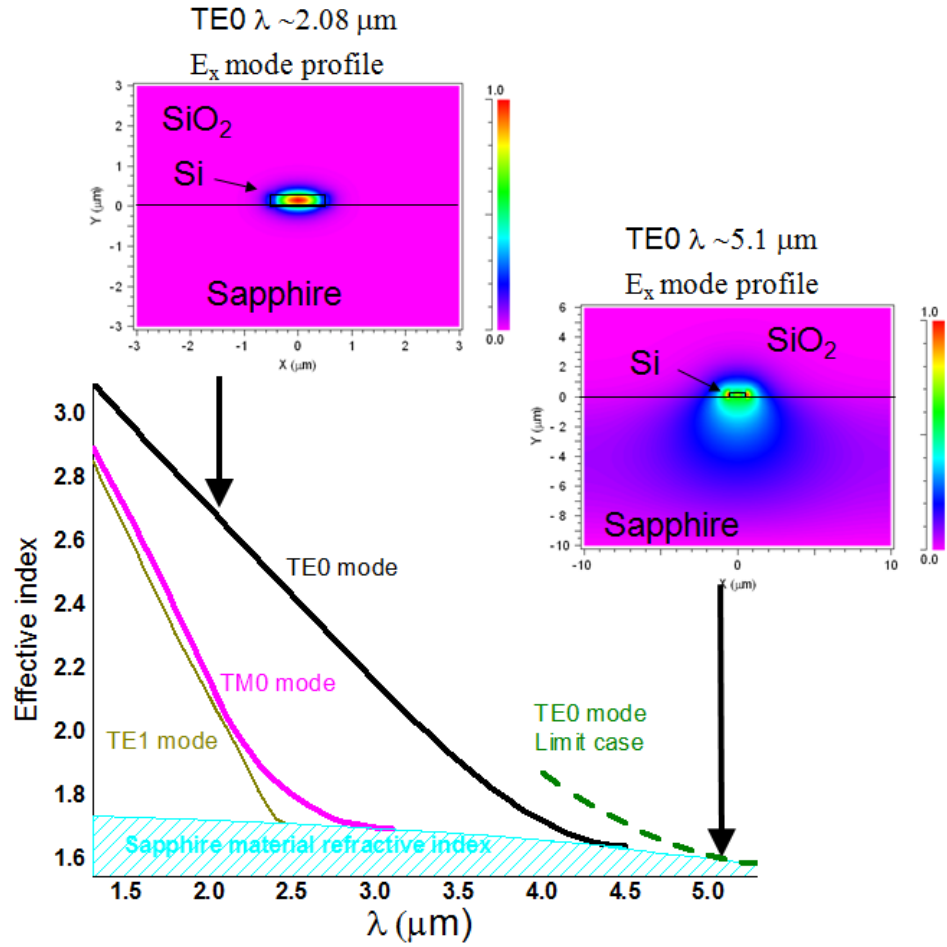

Figure 5. Calculated effective index vs λ for TE and TM polarizations for 1000nm wide SOS nanowires, based on Finite Element Methods (FemSim RSOFT). The mode profiles are for the $E_x$ component. The blue line is the refractive index for sapphire and so the shaded blue region represents effective indices below cutoff.

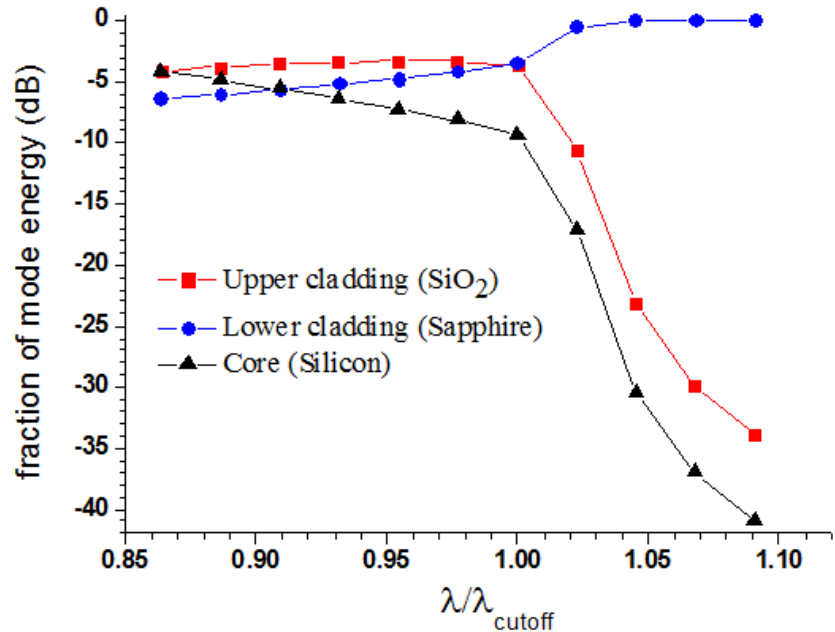

Figure 6. Calculated relative fraction (in dB) of the mode energy in the three different regions (upper cladding, core, lower cladding) as a function of wavelength normalized to cutoff. Also shown is the relative mode area.

## 5. Conclusions

We demonstrate record low propagation loss silicon-on-sapphire nanowires for operation in the mid IR wavelength range, achieving < 1dB/cm at $\lambda$=1.55μm, slightly higher near $\lambda$=2.08μm and < 2dB/cm at $\lambda$=5.18μm.

**Acknowledgements**

This work was supported by the Australian Research Council (ARC) Centres of Excellence program and the ARC Federation Fellows program.